# Cyber Situation Awareness with Active Learning for Intrusion Detection

Steven McElwee
PJM Interconnection
steve.mcelwee@pjm.com

James Cannady
Georgia Tech Research Institute
cannady@gatech.edu

*Abstract*—Intrusion detection has focused primarily on detecting cyberattacks at the event-level. Since there is such a large volume of network data and attacks are minimal, machine learning approaches have focused on improving accuracy and reducing false positives, but this has frequently resulted in overfitting. In addition, the volume of intrusion detection alerts is large and creates fatigue in the human analyst who must review them. This research addresses the problems associated with event-level intrusion detection and the large volumes of intrusion alerts by applying active learning and cyber situation awareness. This paper includes the results of two experiments using the UNSW-NB15 dataset. The first experiment evaluated sampling approaches for querying the oracle, as part of active learning. It then trained a Random Forest classifier using the samples and evaluated its results. The second experiment applied cyber situation awareness by aggregating the detection results of the first experiment and calculating the probability that a computer system was part of a cyberattack. This research showed that moving the perspective of event-level alerts to the probability that a computer system was part of an attack improved the accuracy of detection and reduced the volume of alerts that a human analyst would need to review.

I. INTRODUCTION

Current machine learning approaches to intrusion detection are insufficient for dealing with the problems of uncertain results, overfitting, and large volumes of alerts for human review [1]. At present, intrusion detection focuses on the observation of discrete events to identify cyberattacks, which can be error prone and insensitive to novel attacks. When human analysts respond to intrusion alerts, they use mental models of cyber situation awareness to combine the observations of discrete events and create an understanding of the current situation [2]. Once the current situation is understood, it is possible to project a future state and determine an appropriate course of action. But human analysts rely on uncertain information provided by intrusion alerts and use their own biases, which can result in bad decisions [3]. This problem is compounded by the large volume of events that must be analyzed.

A new approach to intrusion detection is needed to enhance decision-making and enable more rapid response to cyberattacks. This research addresses the problems associated with intrusion detection by applying active learning to support cyber situation awareness. This new approach to intrusion detection applies machine learning to intrusion detection and is supported by probabilistic cyber situation awareness to identify the computer systems involved in an attack. This allows improved decision-making to identify an appropriate course of action. This research demonstrated that high detection accuracy at the event-level was unnecessary to predict that a computer system was carrying out attacks. Instead of over-training detection algorithms, this solution detected novel attacks and was more robust against adversarial evasion.

Using a series of active learning iterations, this research classified the UNSW-NB15 intrusion detection dataset to create multiple sets of observations. After manually labeling as few as 160 records, the results demonstrated that by aggregating the observations to the source IP address, the algorithm had a high degree of accuracy in separating the IP addresses that were normal from those that were carrying out cyberattacks.

This research builds upon prior active learning for intrusion detection research [4] in three ways. First, it introduces a new approach to query selection for active learning. In the previous research, *k*-means clustering was used for sampling. This research introduces *k*-means clustering with bagging for query selection and provides a more detailed comparison of random, *k*-means, and *k*-means with bagging sampling. Second, this research uses a more contemporary dataset for intrusion detection. The prior research utilized the KDD-Cup 1999 dataset, but this research uses the UNSW-NB15 dataset, which better reflects modern network traffic. Third, this research moves from the detection of attacks in individual network events and develops a level of cyber situation awareness. This allows human analysts to rapidly identify the computer systems involved in attacks and to take action on this systems to contain and eradicate the compromises.

The remainder of this paper provides background information on related work in intrusion detection, cyber situation awareness, and active learning. Next, it describes the two experiments conducted for this research and the results of each. After this, it reviews the results, implications, limitations, and potential for future research.

II. RELATED WORK

*A. Intrusion Detection*

Intrusion detection identifies cyberattacks in computer systems and networks [5,6]. Attacks may include unauthorized access to or modification of files, user information, network data, or system resources [7]. Signature-based intrusion detection matches computer or network information to known patterns that represent cyberattacks [8]. Anomaly-based intrusion detection identifies abnormalities in computer systems or network packets to point to potential attacks [8]. Intrusion detection is commonly categorized as either network-based or host-based, although these can be further subdivided [9].

Machine learning approaches are commonly used in intrusion detection. Examples include: symbolist approaches that use decision trees and random forests [4]; connectionist approaches that use neural networks [10-12]; evolutionary approaches that mimic genetics or the immune system [13]; Bayesian methods [14]; and analogistic approaches that use support vector machines [15].

Accuracy is one of the most common methods for evaluating intrusion detection systems. Using a confusion matrix as in Figure 1 accuracy can be defined as all of the true positive and true negative outcomes divided by all possible outcomes:

$$accuracy = \frac{TP + TN}{TP + FP + TN + FN} \qquad (1)$$

Intrusion detection research frequently focuses on false positive reduction [16]. This is especially important because of the highly imbalanced nature of intrusion detection data, since attacks are rare in vast amounts of data [17]. As a result, even a highly accurate intrusion detection system may have high false positive rates [18]. To address this problem, intrusion detection algorithms are commonly over-trained, making them useful only for detecting known attacks and ignoring novel attacks [1].

More important to this research, intrusion detection requires repetitive work for humans to analyze and prioritize alerts, which results in fatigue and bad decisions [3]. Intrusion alerts are generally incomplete, leading to uncertainty, which increases the likelihood of bad decisions. As a result, it is important not only provide alerts for discrete events, but also to develop the context of those alerts to support improved decision-making [19].

*B. Cyber Situation Awareness*

Situation awareness is a theory that was introduced to create a model for how understanding can be developed [20]. By beginning with discrete events or observations, an understanding of the current situation can be developed. Three levels of situation awareness are commonly used, as depicted in Figure 2: 1) observation of elements of the current situation; 2) comprehension of the observations to understand the current situation; and 3) projection of a future state. Although situation awareness has a wide variety of applications, it has been specifically applied to decision-making for security analysts [21].

Theories of cyber situation awareness have been developed using cognitive task analysis to understand the day-to-day practices of security analysts [2,21]. These studies have found that because of the large amounts of data and uncertain information, analysts rely on their own memory of past incidents and make biased decisions. Intrusion detection research has focused on the observation of discrete events, but it has not provided support for moving from uncertainty to help analysts to develop an understanding of the situation [22].

*C. Active Learning*

Active learning is a machine learning approach that focuses on interacting with human experts to label unlabeled data efficiently. The machine learning algorithm selects a sample of records and presents them to the human expert, who is referred to as an oracle [23]. The sample of records is presented to the oracle, which is called a query. A variety of strategies for optimal queries to human experts has been developed to minimize the number of queries to the oracle and still accurately label a dataset. Initial approaches applied set theory, including membership, equivalence, subset, supersets, disjointness, and exhaustiveness [24]. Other approaches have successfully used statistical methods and neural networks [25]. Still others have applied greedy search, opportunistic priors, and Bayesian assumptions [26]. More recent research has found the value in uncertainty sampling [27] as well as using clustering to create a more diverse sample [4].

Active learning has been found to be effective in adversarial environments, since it adapts based on human feedback [27]. This research builds upon prior work that applied active learning

|  | | Predicted Class | |
|---|---|---|---|
|  | | Normal | Attack |
| Actual Class | Normal | True Negative (*TN*) | False Positive (*FP*) |
| | Attack | False Negative (*FN*) | True Positive (*TP*) |

Fig. 1.   Confusion matrix

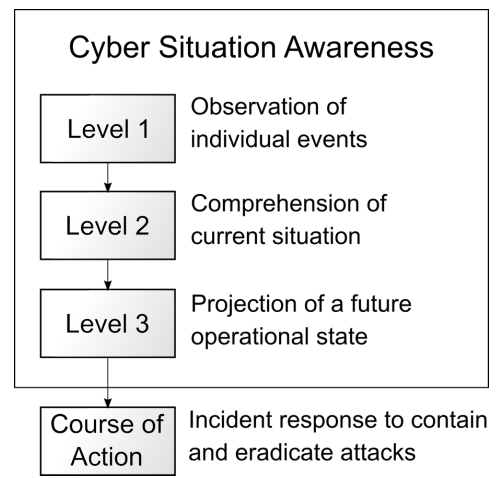

Fig. 2   Levels of cyber situation awareness

to intrusion detection to label a large imbalanced dataset while minimizing human interaction [4].

## III. EXPERIMENTS AND RESULTS

This research conducted of two experiments. The first was to evaluate approaches for sampling the data to find the most effective query for the oracle. The second was to develop cyber situation awareness from event-level detection of cyberattacks to predict the probability that a computer system was part of an attack. The experiments in this research were implemented using Python. This section describes the experiments and their results.

### A. Experiment 1 – Sampling Method Evaluation

#### 1) Design

Using the first dataset of the UNSW-NB15 intrusion data [28], the first experiment sought to find a sampling approach that minimized the number of queries to the oracle. Having a diverse sample that included normal and attack records was important, since the results of this query would then be used to train a Random Forest classifier. The absence of attack records in the sample would make the training of the classifier of little value.

The three approaches for sampling applied were random selection, k-means clustering of all of the features, and k-means clustering with bagging. Given the input records, $X = \{x_1, x_2,...x_n\}$. The goal of the selection was to find a sample, $T$, such that $T \in X$ that maximized diversity by including a mixture of normal and attack records. After selecting the sample, $T$, this set was presented to the Random Forest classifier to label the entire dataset. Thus, $rf(X) = Y$, where $Y$ is the set of labels corresponding to $X$, such that $Y = \{y_1, y_2,... y_n\}$. The labels consisted of $y_n \in \{0, 1\}$, such that 0 represented a normal record and 1 represented an attack.

Figure 3 shows the algorithm used for this experiment. The *number of runs* for this experiment was set to 30. The ActiveLearner class implemented each of the three sampling methods. The while loop was to ensure that none of the samples returned only normal records. For testing the *minimum attack labels* was set to one.

The dataset was loaded into a Pandas DataFrame. The random selection sampling method used the built-in sampling available in DataFrames to select 40 samples. The k-means clustering sampling used all features from the DataFrame and applied the *sklearn k*-means algorithm to create 40 clusters. One random sample was then taken from each cluster. Finally, the k-means clustering with bagging sampling method randomly selected between 20 and 35 features and used the k-means clustering algorithm to create a random number for clusters ranging from 30 to 50.

After selection of the samples using each of the three methods, the samples were used to query the oracle for labeling. Since the UNSW-NB15 dataset contains labels, a human analyst was not required to label the records. Instead this was implemented in software. Using the set of 30 to 50 labeled records provided by the oracle, this experiment proceeded to train a Random Forest classifier. Using the trained classifier, the entire dataset was classified. The results of the classification were used to evaluate each of the sampling methods.

#### 2) Results

Using the labels identified by the Random Forest classifier, the accuracy, sensitivity, and false positive rates were evaluated. Since each approach included a degree of randomness, each of the three sampling methods was tested 30 times to ensure the results were reproduceable. Table I shows a summary of the results. Each sampling method obtained an average accuracy of greater than 97%. This is not surprising, since the UNSW-NB15 dataset is highly imbalanced. What was surprising was that each had a similar sensitivity, or true positive rate of over 30%. This was surprising because it indicated that each method may be similarly capable of identifying a mixture of normal and attack records. Each method also had similar false positive rates of 0.3% and 0.4%.

TABLE I. AVERAGES SAMPLING METHOD RESULTS OVER 10 RUNS

| Sampling | Accuracy (%) | Sensitivity (%) | False Positive Rate (%) |
|---|---|---|---|
| Random | 97.5 | 34.1 | 0.4% |
| k-Means | 97.7 | 37.1 | 0.3% |
| k-Means with Bagging | 97.6 | 32.5 | 0.3% |

Where the sampling methods showed differences was in their variation. Figure 4 shows the distribution of results for accuracy, sensitivity, and false positive rates, respectively. Variations in accuracy for the bagging method provided some results as high as 99.1%. More important, variations in sensitivity for bagging resulted in true positive rates as high as 96.2%. Bagging also showed a larger portion of low false positive rates than random sampling or *k*-means clustering selection. Although any of the three algorithms were found to be suitable for active learning, *k*-means with bagging was selected for the remainder of this research because the wide variation provided a more diverse set of observations for predicting the probability that a computer system was involved in an attack.

Since the bagging method was selected for the remainder of this research, it is important to note its average sensitivity, or

```
Initialize configurable variables
for i = 0 to number of runs do
    Instantiate ActiveLearner
    while count of attacks < minimum attack labels
        Create sample using selected method
        Query the oracle with the sample for labeling
    Train the Random Forest classifier using sample labels
    Classify the entire dataset
    Evaluate results of classification
end for
```

Fig. 3. Algorithm for experiment 1

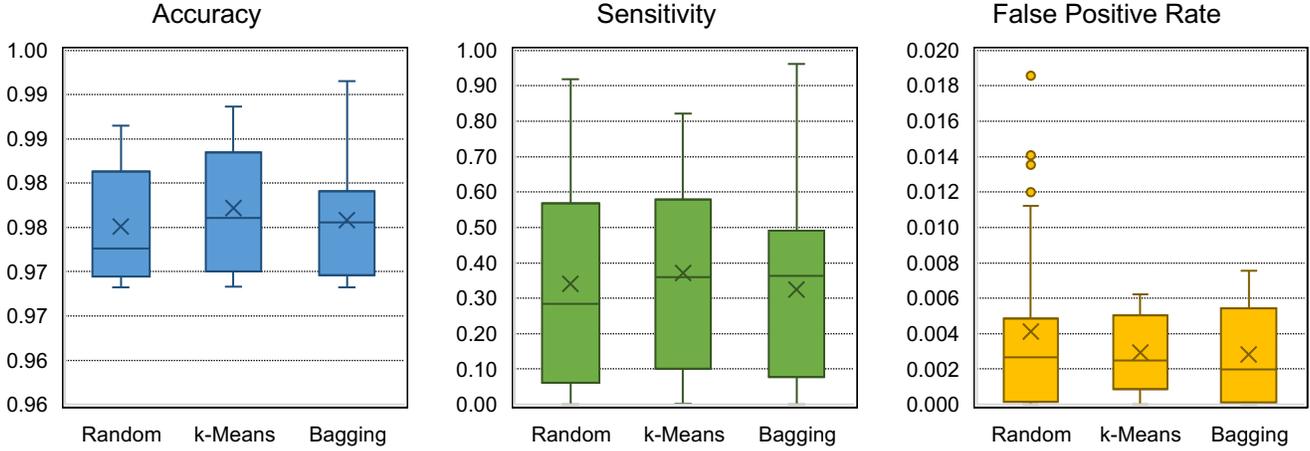

Fig. 4 – Experiment 1 distribution of results

true positive rate, of 32.5%. This will be used as the probability for the Bernoulli trials in the second experiment.

B. *Experiment 2 – Cyber Situation Awareness*

1) *Design*

The second experiment used the observations, which were the detected attacks, from the first experiment to evaluate the probability that a computer system was part of an attack. Bernoulli trials were used to combine the observations to determine if each computer system was involved in an attack. A Bernoulli trial was appropriate for calculating the cumulative probability, since this experiment met the three conditions for using a such a trial. First, there were only two possible outcomes, in that the classifier only predicted normal and attack records. Second, each detection had a fixed probability, $p$, of being a successful detection of a normal attack. This was found in the previous experiment to be the sensitivity of 0.325. By extension, using the miss rate of 1 – sensitivity, yielded $q$ = 0.675. Third, each trial was completely independent of all others. This condition was met, since each run was based on a different active learning sample and the classifier for each run was trained separately.

Using the bagging method from the first experiment, the second experiment ran the active learning sequence 10 times. The active learning sequence included selecting a sample from the dataset, training the classifier using the sample, and classifying the entire dataset as the testing. Each of these 10 sequences and results was considered a "run".

To implement this experiment, the detection results from each run were grouped by *srcip*, which was the feature in the dataset used to represent each computer system. Since each detection result was either a 0 for normal records or a 1 for a detected attack, the grouped results were summed to find the total number of attack outcomes. Thus, for each *srcip* the number of detected attacks, $d$, was:

$$d = \sum Y_d \qquad (2)$$

The total number of trials, $n$, was the number of records, including both normal and attack, associated with each *srcip*. Thus, for each *srcip*, the probability that that computer system was part of an attack was:

$$P(d) = \binom{n}{d} p^d q^{n-d} \qquad (3)$$

To simulate how a human analyst might use this active learning solution, the results of each run built upon each other. As a result, the number of trials for a given run, $r$, was:

$$n_r = \sum_{i=0}^{r} n_i \qquad (4)$$

The number of detections for a given run was similarly:

$$d_r = \sum_{i=0}^{r} d_i \qquad (5)$$

Thus, the results of each run for $P(d)$ reflected all the trials and all the detections leading up to the run. The algorithm to calculate $P(d)$ was the same as the algorithm from the first experiment, but added the summation of $n$ and $d$, as well as the calculation of $P(d)$ for each *srcip* for each run.

2) *Results*

Each run of the active learning cycle provided a variety of detection results, and each had a degree of error. As a result, the first run yielded very low probabilities that any systems were involved in an attack. By the end of the second run, however, three of the four computer systems that were part of the attacks in the dataset were identified with a 100% probability. Of the remaining computer systems in the dataset 29 had probabilities of 0%, while the rest had probabilities of 70% or lower. By the end of the fourth run, all of the attacking computer systems had been identified with 100% probability, and the remaining computer systems had probabilities of less than 50%. The Appendix of this paper shows the results of each run.

Figure 5 graphs the probability of each *srcip* being involved in an attack for each of the runs for only the attacking computers.

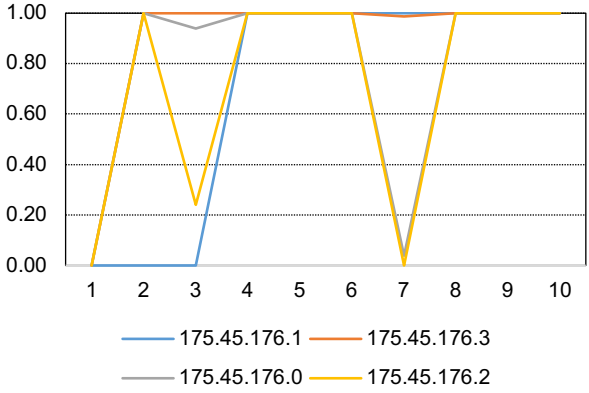

Fig. 5. $P(d)$ for attacking *srcip* computers for rounds 1 through 10

It is interesting to note that run seven detected few attacks and reduced the probabilities of two of the computer systems down to zero or near zero. This was to be expected because of the randomness associated with each run. By the next run, all four computer systems were predicted with 100% probability to be part of an attack. Figure 6 shows a similar view for three of the non-attacking computer systems that experienced a reduction in probability as the runs progressed.

This experiment found that by requesting a human analyst to label an average of 40 records per run, after four runs, with 160 labels manually identified, the computer systems involved in attacks could be identified accurately in a dataset of 700,001 network events. At a more extreme case, if all 10 runs were required, the human analyst would only need to label 400 events. Using the list of computer systems that are involved in an attack, an analyst can move quickly into a course of action to contain and eradicate the compromise.

*C. Source Code and Detailed Results*

The source code used for these experiments as well as the detailed results are available for download at https://github.com/stevenmcelwee/alcsa. This will allow the results to be reproduced and expanded upon in future research.

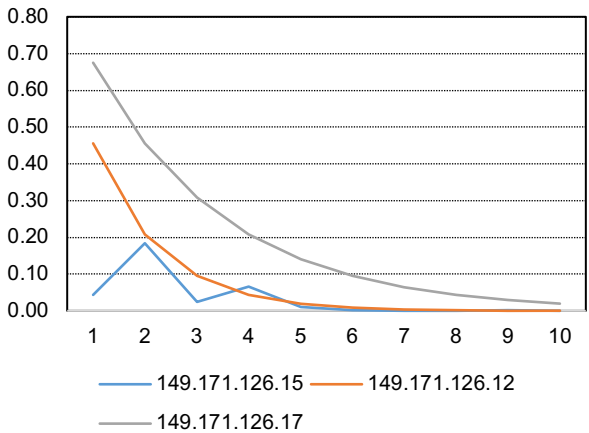

Fig. 6. $P(d)$ for non-attacking *srcip* computers for rounds 1 through 10

## IV. DISCUSSION

This research made four main contributions. First, it built upon a similar previous study [4] by using a more suitable dataset for intrusion detection research, the UNSW-NB15 dataset. This made the research more relevant and practical for application. Second, it provided a more in-depth analysis of the selection of a sampling method for developing queries for the oracle. Although this study determined that the three sampling strategies provided similar results, it also provided additional criteria for evaluation by looking more closely at sensitivity, false positive rates, and variation in the results. Third, this study showed how making a few runs of active learning cycles, the probability that a computer system was involved in an attack could be calculated cumulatively with a high degree of accuracy. Fourth, this research demonstrated the value in moving the perspective of intrusion detection from event-level detection to evaluating the probability that a computer system was part of an attack. By doing so, this research demonstrated that by manually labeling between 160 and 400 records in the dataset, a human analyst could quickly determine the computer systems that require incident response activities. Thus, this solution addressed previous concerns about the volume of intrusion alerts and the fatigue associated with reviewing them.

In addition, this research was novel in that it applied active learning to cyber situation awareness. It moved intrusion detection from the first level of situation awareness into level two, by aggregating alerts with cumulative probability. As a result, this research set the stage for future cyber situation awareness research that can further apply machine learning to predicting a future operational state and even taking a course of action.

This research had three known limitations. First, because it used a single dataset for intrusion detection research, it may not apply to all intrusion datasets or to real network traffic. This research used an accepted dataset to reduce the impact of this limitation, but testing was not performed on live operational networks.

Second, this research did not account for the limitations of human experts to provide the correct labels in the active learning cycle. Instead, the oracle in these experiments provided the correct label 100% of the time. The impact of this limitation is reduced because the results are used to train a Random Forest classifier, which introduced some error. This classifier was chosen because it is less prone to overfitting. As a result, it was expected that a noisy oracle would not introduce a significant change in these results.

Third, this research was limited in that the solution only supports off-line analysis. Although the algorithms used in this research ran quickly on the test dataset, real operational network data is significantly more voluminous and takes longer to process. As a result, the results of this research are best suited for off-line analysis and incident response, rather than early detection of attacks in progress.

## V. CONCLUSION

Intrusion detection research has focused primarily on event-level detection, which experiences problems because of large volumes of data, highly imbalanced datasets, and human fatigue.

This research addressed these problems by using cyber situation awareness to shift the focus from the event-level to the computer system level. The experiments in this research showed that active learning using *k*-means clustering with bagging was effective in providing a diverse sample to the oracle. They also showed that by aggregating the event-level detection results to the computer system level, a Bernoulli trial was effective in calculating the probability that a computer system was part of an attack. The results showed that with minimal human interaction to label the samples, the solution provided highly accurate results that could allow a human analyst to more quickly comprehend the current situation and to take action to contain and eradicate cyberattacks.

APPENDIX – RESULTS OF EXPERIMENT TWO PROBABILITY BY SRCIP

| srcip | Type | n | Run, P(d) | | | | | | | | | |
|---|---|---|---|---|---|---|---|---|---|---|---|---|
| | | | 1 | 2 | 3 | 4 | 5 | 6 | 7 | 8 | 9 | 10 |
| 59.166.0.2 | normal | 67,209 | 0.000 | 0.000 | 0.000 | 0.000 | 0.000 | 0.000 | 0.000 | 0.000 | 0.000 | 0.000 |
| 59.166.0.0 | normal | 67,128 | 0.000 | 0.000 | 0.000 | 0.000 | 0.000 | 0.000 | 0.000 | 0.000 | 0.000 | 0.000 |
| 59.166.0.5 | normal | 67,091 | 0.000 | 0.000 | 0.000 | 0.000 | 0.000 | 0.000 | 0.000 | 0.000 | 0.000 | 0.000 |
| 59.166.0.4 | normal | 66,722 | 0.000 | 0.000 | 0.000 | 0.000 | 0.000 | 0.000 | 0.000 | 0.000 | 0.000 | 0.000 |
| 59.166.0.1 | normal | 66,587 | 0.000 | 0.000 | 0.000 | 0.000 | 0.000 | 0.000 | 0.000 | 0.000 | 0.000 | 0.000 |
| 59.166.0.3 | normal | 66,145 | 0.000 | 0.000 | 0.000 | 0.000 | 0.000 | 0.000 | 0.000 | 0.000 | 0.000 | 0.000 |
| 59.166.0.6 | normal | 64,689 | 0.000 | 0.000 | 0.000 | 0.000 | 0.000 | 0.000 | 0.000 | 0.000 | 0.000 | 0.000 |
| 59.166.0.8 | normal | 64,640 | 0.000 | 0.000 | 0.000 | 0.000 | 0.000 | 0.000 | 0.000 | 0.000 | 0.000 | 0.000 |
| 59.166.0.9 | normal | 64,187 | 0.000 | 0.000 | 0.000 | 0.000 | 0.000 | 0.000 | 0.000 | 0.000 | 0.000 | 0.000 |
| 59.166.0.7 | normal | 63,725 | 0.000 | 0.000 | 0.000 | 0.000 | 0.000 | 0.000 | 0.000 | 0.000 | 0.000 | 0.000 |
| **175.45.176.1** | **attack** | **14,325** | **0.000** | **0.000** | **0.000** | **1.000** | **1.000** | **1.000** | **1.000** | **1.000** | **1.000** | **1.000** |
| 149.171.126.18 | normal | 6,010 | 0.000 | 0.000 | 0.000 | 0.000 | 0.000 | 0.000 | 0.000 | 0.000 | 0.000 | 0.000 |
| **175.45.176.3** | **attack** | **5,128** | **0.000** | **1.000** | **0.999** | **1.000** | **1.000** | **1.000** | **0.988** | **1.000** | **1.000** | **1.000** |
| **175.45.176.0** | **attack** | **4,782** | **0.000** | **1.000** | **0.939** | **1.000** | **1.000** | **1.000** | **0.038** | **1.000** | **1.000** | **1.000** |
| **175.45.176.2** | **attack** | **3,236** | **0.000** | **1.000** | **0.240** | **1.000** | **1.000** | **1.000** | **0.000** | **1.000** | **1.000** | **1.000** |
| 10.40.85.1 | normal | 1,680 | 0.000 | 0.000 | 0.000 | 0.000 | 0.000 | 0.000 | 0.000 | 0.000 | 0.000 | 0.000 |
| 10.40.182.1 | normal | 1,670 | 0.000 | 0.000 | 0.000 | 0.000 | 0.000 | 0.000 | 0.000 | 0.000 | 0.000 | 0.000 |
| 10.40.85.30 | normal | 888 | 0.000 | 0.000 | 0.000 | 0.000 | 0.000 | 0.000 | 0.000 | 0.000 | 0.000 | 0.000 |
| 10.40.170.2 | normal | 874 | 0.000 | 0.000 | 0.000 | 0.000 | 0.000 | 0.000 | 0.000 | 0.000 | 0.000 | 0.000 |
| 10.40.182.3 | normal | 874 | 0.000 | 0.000 | 0.000 | 0.000 | 0.000 | 0.000 | 0.000 | 0.000 | 0.000 | 0.000 |
| 149.171.126.1 | normal | 251 | 0.000 | 0.000 | 0.000 | 0.000 | 0.000 | 0.000 | 0.000 | 0.000 | 0.000 | 0.000 |
| 149.171.126.5 | normal | 249 | 0.000 | 0.000 | 0.000 | 0.000 | 0.000 | 0.000 | 0.000 | 0.000 | 0.000 | 0.000 |
| 149.171.126.6 | normal | 241 | 0.000 | 0.000 | 0.000 | 0.000 | 0.000 | 0.000 | 0.000 | 0.000 | 0.000 | 0.000 |
| 149.171.126.2 | normal | 232 | 0.000 | 0.000 | 0.000 | 0.000 | 0.000 | 0.000 | 0.000 | 0.000 | 0.000 | 0.000 |
| 149.171.126.3 | normal | 231 | 0.000 | 0.000 | 0.000 | 0.000 | 0.000 | 0.000 | 0.000 | 0.000 | 0.000 | 0.000 |
| 149.171.126.4 | normal | 225 | 0.000 | 0.000 | 0.000 | 0.000 | 0.000 | 0.000 | 0.000 | 0.000 | 0.000 | 0.000 |
| 149.171.126.9 | normal | 217 | 0.000 | 0.000 | 0.000 | 0.000 | 0.000 | 0.000 | 0.000 | 0.000 | 0.000 | 0.000 |
| 149.171.126.8 | normal | 217 | 0.000 | 0.000 | 0.000 | 0.000 | 0.000 | 0.000 | 0.000 | 0.000 | 0.000 | 0.000 |
| 149.171.126.7 | normal | 207 | 0.000 | 0.000 | 0.000 | 0.000 | 0.000 | 0.000 | 0.000 | 0.000 | 0.000 | 0.000 |
| 149.171.126.0 | normal | 192 | 0.000 | 0.000 | 0.000 | 0.000 | 0.000 | 0.000 | 0.000 | 0.000 | 0.000 | 0.000 |
| 192.168.241.243 | normal | 108 | 0.000 | 0.000 | 0.000 | 0.000 | 0.000 | 0.000 | 0.000 | 0.000 | 0.000 | 0.000 |
| 149.171.126.11 | normal | 16 | 0.002 | 0.000 | 0.000 | 0.000 | 0.000 | 0.000 | 0.000 | 0.000 | 0.000 | 0.000 |
| 149.171.126.15 | normal | 8 | 0.043 | 0.184 | 0.024 | 0.066 | 0.011 | 0.001 | 0.000 | 0.001 | 0.002 | 0.000 |
| 149.171.126.16 | normal | 6 | 0.095 | 0.009 | 0.001 | 0.000 | 0.000 | 0.000 | 0.000 | 0.000 | 0.000 | 0.000 |
| 149.171.126.19 | normal | 3 | 0.308 | 0.095 | 0.029 | 0.009 | 0.003 | 0.001 | 0.000 | 0.000 | 0.000 | 0.000 |
| 149.171.126.10 | normal | 3 | 0.308 | 0.697 | 0.398 | 0.417 | 0.431 | 0.442 | 0.275 | 0.292 | 0.465 | 0.320 |
| 149.171.126.12 | normal | 2 | 0.456 | 0.208 | 0.095 | 0.043 | 0.020 | 0.009 | 0.004 | 0.002 | 0.001 | 0.000 |
| 127.0.0.1 | normal | 1 | 0.675 | 0.456 | 0.308 | 0.208 | 0.140 | 0.095 | 0.064 | 0.043 | 0.029 | 0.020 |
| 149.171.126.13 | normal | 1 | 0.675 | 0.456 | 0.308 | 0.208 | 0.140 | 0.095 | 0.064 | 0.043 | 0.029 | 0.020 |
| 149.171.126.17 | normal | 1 | 0.675 | 0.456 | 0.308 | 0.208 | 0.140 | 0.095 | 0.064 | 0.043 | 0.029 | 0.020 |